\documentstyle[preprint,aps]{revtex} 
\begin{document} \draft

\title{ OPTICAL RESPONSE OF GRATING-COUPLER-INDUCED INTERSUBBAND
RESONANCES: THE ROLE OF WOOD'S ANOMALIES}

\author{\sc L.  Wendler \\ {\it Anna-Siemsen-Stra{\ss}e 66, D-07745
Jena,
Germany} \\ T.  Kraft, M. Hartung, A. Berger, and A. Wixforth \\
{\it Sektion Physik, Ludwigs-Maximilians-Universit\"at M\"unchen,
Geschwister-Scholl-Platz 1,
D-80539 M\"unchen, Germany}
\\ and \\  M. Sundaram, J.H. English, and A.C. Gossard \\
{\it Materials Department, University of California,
Santa Barbara, CA 93106, U.S.A.}}

\date{\today}

\maketitle

\begin{abstract}
Grating-coupler-induced collective intersubband transitions in a
quasi-two-dimensional electron system
are investigated both experimentally and
theoretically. Far-infrared transmission experiments are performed on
samples containing a quasi-two-dimensional electron gas quantum-confined
in a parabolic quantum well. For rectangular shaped grating
couplers of different periods we observe a strong dependence of the
transmission line shape and peak height on the period of the grating,
i.e. on the wave
vector transfer from the diffracted beams to the collective intersubband
resonance. It is shown that the line shape transforms with increasing
grating period from a Lorentzian into a strongly asymmetric line shape.
Theoretically, we treat the problem by using the transfer-matrix
method of local optics and apply the modal-expansion method to calculate
the influence of the grating. The optically uniaxial
quasi-two-dimensional electron gas is described in the
long-wavelength limit of the random-phase
approximation by a local dielectric tensor, which includes size
quantization
effects. Our theory reproduces excellently the experimental line shapes.
The deformation of the transmission line shapes we explain by the
occurrence
of both types of Wood's anomalies.
\end{abstract}

\bigskip


\section{Introduction}

The optical and electronic properties of electron systems with reduced
dimensionality like {\it quasi-two-dimensional electron gases} (Q2DEG's)
formed at, e.g., a modulation doped semiconductor heterojunction have
been widely studied in the recent past. Especially the collective
excitation spectrum of such a system has attracted a lot of attention as
it represents one of the Q2DEG's most fundamental properties \cite{tk1},
but also
because of potential device applications \cite{tk2}-\cite{tk4}.
Here, both types of collective
charge-density excitations, {\it intrasubband} as well as
{\it intersubband
plasmons} in the {\it far-infrared}
(FIR) regime have been intensively studied for the last twenty years
(see e.g. Ref. \cite{tk5}).

It is known (see e.g. \cite{tk34}) that the intrasubband plasmons
always exist
for wave vectors ${\bf q_\parallel}$
larger than the wave vector of the freely propagating light,
${q_\parallel > \omega/c}$, where ${\bf q}_\parallel=
(q_x, q_y)$ is the in-plane wave vector
(${q_\parallel=|{\bf q}_\parallel|}$)
of the collective excitation (c: vacuum speed of light, $\omega$:
(angular) frequency), assuming the interfaces of the sample to be
parallel to the x-y plane. But differently, each intersubband excitation
is accompanied by two branches of dispersion curves due to the
{\it polariton
effect} \cite{tk6}. One branch is located to the left of the light line
 ${(q_{\parallel}
< \omega / c)}$ and thus, is the dispersion relation of a {\it radiative
virtual mode}, and the second appearing to the right of the light line
${q_{\parallel}> \omega /c}$ describes a
{\it non-radiative normal mode}.
Commonly, one calls
the Q2D radiative virtual intersubband modes the {\it collective
intersubband
resonances} (ISR's), collective intersubband transitions or dimensional
resonances and the non-radiative normal modes the intersubband plasmons.

Unfortunately, the investigation of the mode dispersion of the
Q2D plasmons ${\omega_p^{Q2D}
(q_\parallel)}$ is not directly accessible in FIR spectroscopy. In
general, the Q2D plasmons are accompanied by different collective
intra- and intersubband transitions due to the {\it intersubband
coupling} (ISC). In this case, the resulting spectrum is of {\it hybrid
type}. Only for weak ISC nearly pure intra- and intersubband plasmons
occur. In such a situation,
intrasubband plasmons can only be excited with fields polarized parallel
to the heterointerfaces and which have wave vectors ${q_{\parallel}
> \omega / c}$, whereas ISR's and intersubband plasmons can only be
excited with electromagnetic fields having components
polarized perpendicular to the interfaces of the
sample \cite{tk7,tk8} and which have wave vectors
${q_{\parallel} < \omega / c}$ and
${q_{\parallel} > \omega / c}$, respectively.
To solve this problem, usually the FIR
radiation is coupled to the Q2DEG through a metallic
grating of period $d$ above the electron system \cite {tk8}
-\cite{tk10}. This way,
different discrete values of the probe wave vector
${\bf k}_{\parallel n}=(k_{xn}, k_y)$ of the diffracted
electromagnetic field parallel to the 2D plane,
$k_{xn}=(\omega /c) \sin \Theta _0 + (2 \pi  /d) n ; n=0, \pm 1, \pm 2,
\ldots$ ($\Theta_0$: ray angle of the incident light
measured from the $z$-axis, which is assumed
to be perpendicular to the interfaces of the sample),
can be excited. Assuming the incident light propagation in the x-z
plane and the stripes of the grating along the y axis, the incident
light couples at the discrete
wave vectors $ q_\parallel = k_{xn}$ to the collective
excitations $\omega_p^{Q2D}(q_\parallel)$ in the Q2DEG, provided
the grating is sufficiently close to the electron gas.
Here, the grating coupler serves as a
special optical coupling arrangement: ($i$) to convert the in-plane
electromagnetic
field component  $E_x(x,z|\omega)$ of the incident wave into
the perpendicular component $E_z(x,z|\omega)$
of the scattered waves and ($ii$) induces a wave vector
transfer $k_x \rightarrow
k_{xn}$ parallel to the interfaces of the system. Moreover, grating
couplers provide unique advantages over other techniques like the use
of prisms \cite {tk11}, Brewster angle orientation \cite {tk12},
or multiple internal
reflection (wave guide) geometry \cite {tk13}: normal incidence of the
FIR radiation and a relatively large
component of the electric field along the
direction of confinement can be easily achieved.

The aim of this paper is to investigate both experimentally and
theoretically the influence of the grating coupler on the coupling
efficiency
and the resulting transmission line shape of the Q2D ISR as a
function of the grating coupler period. Both, intrasubband plasmons
(see e.g. Refs. \cite{tk8}-\cite{tk10} for experiments and
Refs. \cite{tk20}-\cite{tk24} for theory) and ISR's
(see e.g. Refs. \cite{tk14}-\cite{tk17} for earlier experiments and
Refs. \cite{tk18,tk19} for theory) are intensively investigated.
Here, we wish to focus on the excitation of the ISR.
Unfortunately, the up to now developed theories
on the {\it grating-coupler-induced} excitation of collective
Q2D plasmon modes is
restricted to strong approximations, e.g. optically isotropic media,
2D gratings, perfectly conducting gratings, simple half-space
geometry etc.,
which are unsuitable to describe correctly the optical response of the
Q2DEG synthesized in multilayer systems with grating.
Moreover, as we will point out below, the influence of the grating
coupler on the line shape is much more pronounced for the ISR than for
the intrasubband plasmon and thus,
needs to be considered carefully. Because the perpendicular field
component is zero for the zeroth-order beam, the grating-coupler-induced
ISR's are nonvertical transitions in ${\bf k}$-space, i.e. involve
wave vector transfers of $k_{xn}, n= \pm 1, \pm 2, \ldots$.
Experiments \cite{tk24a,tk24b} indicate that the power absorbed at the
ISR frequency is a function of the ratio of the grating period $d$ to
the intersubband resonance wavelength $\lambda_{sub}$.
Our theoretical investigations are
completely universal in the framework of local optics and can be applied
to any given optically uniaxial multilayered structure with gratings of
finite height.

\section{Experiment}

The experiments are performed on {\it parabolic quantum well's} (PQW's)
\cite {tk25}.
Here, the conduction band edge of an $GaAs-Ga_{1-x}Al_xAs$ quantum well
is graded such that it turns out to
have a parabolic shape in the growth direction. This is achieved by a
proper variation of the percentage of one of the ingredients (namely the
Al-content x). An electron which is experiencing this profile 'sees' a
potential which could also result from a homogenous positive
background such as a doping layer. From Poisson's equation and a
back-of-the-envelope calculation the potential of this (in the case of a
PQW
{\it fictitious})  positive background with density $n_+$  can be
related to the growth parameters of the PQW by
\begin{equation}
n_+ = \frac{\varepsilon_0\varepsilon_s}{e^2}\frac{d^2E_C}{dz^2}=\frac{
\varepsilon_0\varepsilon_s}{e^2} \frac{8 \Delta }{W^2}.
\label{1}
\end{equation}
Here,  $  \varepsilon {}_{ s}{  }$ denotes the mean static
dielectric constant of
$Ga_{1-x}Al_xAs$ forming the PQW,  $ \varepsilon {}_{0}$ is the
dielectric permittivity in vacuum,
$  \Delta {} $  is the energy height of the parabola from its bottom to
the edges, the electron charge is -e and W is the width of the grown PQW.
Once such a  structure is remotely doped, the donors release electrons
into the well which in turn will screen the man-made parabolic potential
and form a wide and nearly homogeneous electron layer.

The collective response of such systems has been the subject of many
different experiments and theoretical investigations over the last few
years (see e.g. Ref. \cite {tk26}). The most surprising effect
that has been observed in the
early days of the spectroscopy of PQW's is that the electron system
absorbs radiation independently from the electron density in the PQW
only at a single well defined frequency,
which can be related to the shape of the bare confining potential alone.
Self-consistent effects like the renormalization of the subband
energies in the
structures due to electron-electron interactions seems to turn
out to be of no
importance for the optical absorption. In a celebrated paper
Brey, Johnson and Halperin \cite{tk27} and Yip \cite{tk28}
generalized Kohn's theorem \cite{tk29}, originally derived for the
cyclotron resonance in 3D bulk semiconductor structures. They showed
that for
the case of a {\it bare perfectly parabolic confinement potential}
long wavelength
radiation only couples to the {\it center-of-mass} (CM) part of the
Hamiltonian, leaving the {\it relative coordinates} completely
untouched. The
reason for this unique phenomenon is that for a bare harmonic
confinement
and only for this the Hamiltonian can be separated into one part
containing only CM coordinates and one part containing only relative
coordinates. The interaction Hamiltonian  for the incident light can be
shown to couple only to the CM part. In other words, long wavelength
radiation excites a mode in the electron system, which is only connected
with the CM motion. This collective intersubband mode is sometimes
referred to as {\it Kohn's mode} or more figurative
'{\it sloshing mode}'.
{\it Per construction}, it corresponds to the classical plasma
oscillations of a bulk 3D electron gas of density
$n_{+}: \omega_p=(n_+e^2/(\varepsilon_0 \varepsilon_sm_e))$, where
$m_e$ is the effective conduction-band-edge mass.
If we use Eq. (\ref{1}) relating
$n_{+}$ to the growth parameters it reads:
\begin{equation}
\omega_p=
\Omega =
\left(\frac{8\Delta}{W^2m_e}\right)^{\frac{1}{2}},
\label{2}
\end{equation}
where $\Omega$ is the confining frequency of the bare parabolic
potential: $V_0(z)={1\over 2}m_e \Omega^2 z^2$.

The many-particle picture of this phenomenon \cite{tk30,tk31}
is that the {\it renormalization} of the (1-0) subband
separation frequency $ \Delta \Omega_{10}= \Omega_{10}(0)- \Omega_{10}
(n_{2DEG})$, where
$\Omega_{10}(n_{2DEG})=({\cal E} _1 -{\cal E} _0)/ \hbar $ is the
subband separation frequency of the PQW with the 2D electron density
$n_{2DEG}$
(sheet carrier concentration), cancels for $q_\parallel=0$
with the {\it collective frequency shift}
$\Delta _p^{10}= \omega _p^{10}(q_{\parallel} =0)- \Omega_{10}
(n_{2DEG})$
of the (1-0) intersubband plasmon $\omega_p^{10}(q_{\parallel})$.
It is shown \cite{tk30,tk31} that in such a situation this mode
is pinned at  $\omega_p^{10} (q_{\parallel} =0)= \Omega_{10}(0)= \Omega$
independently from the electron density. Whereas Kohn's theorem
states that the absorption spectrum has only one peak at
$\omega=\Omega$, the mode spectrum of the freely oscillating Q2DEG
consists of all types of Q2D plasmons. It should be noted that for
$ q_{\parallel} \ne 0$ Kohn's theorem is not valid.

For our experiments we
use different samples that have been cut from one single wafer. On top
of these samples we fabricate metal gratings of different period $d$ and
then compare their transmission spectra. Any differences in the spectra
obtained for the different  samples thus can be related to the effect of
the different grating couplers.
The sample used in our experiments is a standard PQW
(sample PB48 of Ref. \cite{tk26}), schematically drawn in Fig.1,
in which the Al content x was varied during growth between
$x=0$  in the center of the well and $x =0.3$  at its edges.
To grade the structure, we used the digital alloy technique as
described elsewhere \cite{tk25}.
This structure is grown on a semi-insulating GaAs substrate and consist
of the following layers: layer 1 is the 10 nm thick undoped $GaAs$ cap
layer ($c_0$),
layer 2 is a 200 nm thick undoped $Ga_{0.7}Al_{0.3}As$ layer ($c_1$),
layer 3 is a  17 nm thick Si doped ($N_d = 5 \times 10^{17} cm^{-3}$)
$Ga_{0.7}Al_{0.3}As$ layer ($d_0$), layer 4 is a 4nm thick Si doped
($N_d = 1.11 \times 10^{18} cm^{-3}$) $Ga_{0.7}Al_{0.3}As$ layer ($d_1$),
layer 5
is the 20 nm thick undoped $Ga_{0.7}Al_{0.3}As$ spacer layer ($s_1$),
layer 6
is the 130 nm thick $Ga_{1-x}Al_{x}As$ PQW ($L$), layer 7
is the 20 nm thick undoped $Ga_{0.7}Al_{0.3}As$ spacer layer ($s_2$),
layer 9 is a 4nm thick Si doped ($N_d = 1.11 \times 10^{18} cm^{-3}$)
$Ga_{0.7}Al_{0.3}As$ layer ($d_2$),
layer 10 is a 200 nm thick undoped $Ga_{0.7}Al_{0.3}As$ layer ($c_2$) and
layer 11 is the 500 nm thick undoped $GaAs$ buffer layer ($b$).  The
substrate is a 500$\mu$m thick GaAs wafer. From
the growth parameters we calculate the density of the fictitious charge
of $n_{+}=7.4 \times 10 ^{16} cm^{ -3}$  which according to Eq. (\ref{2})
corresponds to an expected resonance energy of about $\hbar \omega
_p^{10}(q_\parallel=0) \approx 11 meV$. A semi-transparent
NiCr electrode on top of the sample serves as a gate to vary
the carrier density in the well and alloyed In pellets at its corners
provide Ohmic contacts to the electron system. On top of the NiCr gate
a 50nm thick rectangular Ag grating with stripes along the y axis
of width $a$, spacing
$b$ between the stripes and period $d=a+b$ is deposited. For the different
samples investigated here, we used grating periods of
$d=$  4, 6, 10, 20, 27, 40 and 80  $  \mu{}$m, respectively. The
metallization for all gratings was chosen to be in such a way
that the mark-to-space
ratio (aspect ratio) is close to $t=a/b=1$
(one also introduces the mark fraction
$f=a/d$ and the open space fraction $r=b/d$).
For comparison, also one sample without any grating coupler has been
fabricated. To avoid Fabry-Perot type interference effects, the sample
substrate is wedged by a small angle of approximately 3 degrees.

Magneto-transport measurements revealed a typical sheet carrier
concentration of
$n_{ 2DEG}{  } =2 \times 10^{11}cm^{-2}$  for all samples. To
be able to directly compare the experimental results for different
samples we made sure that during the actual measurement of the FIR
transmission for all samples
$n_{ 2DEG }$  was exactly the same by eventually applying a small
correction bias to the gate electrode \cite{tk32}.

Our transmission experiments are performed at low temperatures (
$T=2K)$
with the sample mounted in the center of a superconducting solenoid.
Experimentally, we determine the relative change in transmission
$ - \Delta {} T/T = [T(0)-T(n_{ 2DEG})] / T(0)$.  T(0)  is the
transmission of the sample with the well being completely depleted,
$T(n_{ 2DEG} )$  the one at finite carrier densities. The spectra have
been taken with a Fourier transform spectrometer and a Si composite
bolometer was used to detect the transmitted radiation which is guided
by a ten inches long oversized brass waveguide between the sample and
the bolometer.

In Fig. 2 we depict the results of our experiments, where we plot the
relative transmission
$ - \Delta {} T/T$  for seven samples, investigated as a function of the
frequency for zero magnetic field. As expected from the dipole
selection rules, there is no detectable absorption at the frequency of
the ISR
for the sample without grating. With increasing grating period a
peak develops at the expected resonance frequency of the ISR. Its line
shape
and peak height, however, strongly depend on the grating used in
the experiment. For small grating periods
$d <  \lambda {}_{ sub} $, where
$\lambda _{sub}=2\pi c / (\sqrt \varepsilon _s
\omega)$ is the wavelength of the FIR radiation at the ISR frequency
in the GaAs substrate with
static dielectric constant $\varepsilon _s $,
the peak has a Lorentzian shape and its height
increases with increasing period.
For $ d=\lambda {}_{ sub} ,$  however, the line shape becomes totally
destorted indicating some interference effect. For even larger grating
periods
$d >  \lambda {}_{ sub }$ the relative transmission rapidly decreases.
For
$d=40  \mu{}$m, the resonance is barely detectable and for a grating
with period
$d=80  \mu{}m,$  no detectable resonance is left. It is also
interesting to note that the sign of the resonant structure changes
when the grating
period crosses the condition
$d = \lambda {} _{ sub}.$  Whereas for $d<  \lambda {}_{ sub}$  the
maximum of
$ - \Delta {}T/T$  is positive, for  $d>  \lambda {}_{ sub }$ the
maximum change in transmission becomes negative, i.e. it becomes a
minimum.

\section{Theory}

For the theoretical investigation of the FIR transmission spectra of
the PQW structure with grating coupler we use our theory, developed in Ref.
\cite {tk33}, which is based on the {\it transfer-matrix
method} of local optics for anisotropic media and apply the
{\it modal-expansion method} to calculate the
influence of the grating. Any details of the theory may be found in
this paper.
The combination of both methods results in a generally
computationally efficient and stable formalism of the optical response
of
multilayer systems with grating \cite {tk33}. As shown in Fig. 3 the
PQW under consideration is modelled by a six-layer system
$\nu=1, \ldots, 6$. Each layer is, in general, characterized
by its dielectric tensor $\varepsilon_{\alpha,\beta}^{(\nu)}(\vec
x,\omega)$,
where $\alpha,\beta=x,y,z$, and by its thickness $d_\nu=|z_\nu-
z_{\nu-1}|$.
The layer $z_1<z<z_0$ contains the rectangular-groove grating of
height
$h\equiv d_1$ and
periodicity $d=a+b$. In the grating
region we
have for the filled stripes $\varepsilon_{\alpha\beta}^{(1)}(\vec
x,\omega)=\varepsilon_\xi(\omega)
\delta_{\alpha\beta}$, where $\xi=a$ if $md<x<md+a$ and $\xi=b$
if $md+a<x<(m+1)d$, with $m=0,\pm 1,\pm 2,\ldots$.
In the experiments samples with metal gratings
are used. In this case we use the local Drude dielectric function:
$\varepsilon_a(\omega)=1-\omega_{p_a}^2/\omega(\omega + i \gamma_a)$
for the metallic
stripes, where $\omega_{p_a}$
is the plasma frequency and $\gamma_a$ is the phenomenological
damping constant, and we assume for the spacing between the
stripes $\varepsilon_b=1$.
The semiconductor layers of the quantum-well structure are described
by two
different types of dielectric tensors: (i) $\varepsilon_{\alpha
\beta}^{(\nu)}
(\vec x,\omega)=\varepsilon_{\nu}(\omega)\delta_{\alpha\beta}$  in
$z_\nu<z<z_{\nu-1}$,
if the $\nu$th layer is filled by an isotropic medium and (ii)
$\varepsilon_{\alpha\beta}^{(\nu)}(\vec x,\omega)=
\varepsilon_{\alpha\beta}^{(\nu)}(\omega)\delta_{\alpha\beta}$ is a
diagonal tensor if this layer forms the QW and contains the Q2DEG which
behaves optically uniaxial. This is true in the absence of an
external magnetic field, where we have $\varepsilon_{xx}^{(\nu)}=
\varepsilon_{yy}^{(\nu)}\ne 0$, $\varepsilon_{zz}^{(\nu)} \ne 0$ but
$\varepsilon_{xy}^{(\nu)}=\varepsilon_{yx}^{(\nu)}=\varepsilon_{x
z}^{(\nu)}=
\varepsilon_{zx}^{(\nu)}=\varepsilon_{yz}^{(\nu)}=\varepsilon_{zy}
^{(\nu)}=0$.
For the dielectric properties of the background of the $GaAs$ and
$Ga_{1-x}Al_xAs$
layers we use
the so-called $\varepsilon_s$-approximation:
$\varepsilon_\nu (\omega)=\varepsilon_{s\nu}$,
i. e. we neglect the dynamical properties
of the optical phonons, but include their screening by the static
dielectric constant $\varepsilon_{s\nu}$. This is true because
the frequencies of the optical phonons are well above the frequency
of the ISR in the studied PQW's.
That layer which contains the
Q2DEG we describe by a macroscopic local dielectric tensor, which
includes the size-quantization effects on the electrons
in the PQW. The dielectric tensor
of the Q2DEG (non-local) is derived in the framework of the
{\it random-phase approximation} (RPA) of the current-response scheme
in Ref. \cite{tk34}, wherefrom the nonvanishing components
in the optical limit
($q_{\parallel} \rightarrow 0$) follow in the form:
\begin{eqnarray}
&&\varepsilon_{xx}^{(\nu)}(\omega)=\varepsilon_{yy}^{(\nu)}(\omega)=
\varepsilon_{s\nu}(1-
{\omega_0^2 \over \omega(\omega+{i \over \tau_\parallel})}),
\label{3} \\
&&\varepsilon_{zz}^{(\nu)}(\omega)=\varepsilon_{s\nu}\bigl (1-
{\omega_0^2 f_{10}\over \omega^2-\Omega_{10}^2+{i \over
\tau_\perp}\omega }
\bigr ),
\label{4}
\end{eqnarray}
where
\begin{equation}
f_{10}={2m_e\Omega_{10}\over\hbar} z_{10}^2 \label{5}
\end{equation}
with
\begin{equation}
z_{10}=\int_0^{a_{2DEG}} dz \varphi^*_1 (z)\,z
\,\varphi_0(z).  \label{6}
\end{equation}

Herein, $\varphi_K (z)$, $K=0, 1, 2, \ldots ,$ is the envelope wave
function of the PQW and we have defined the plasma frequency by
$\omega_0=[n_{2DEG}e^2/(\varepsilon_0\varepsilon_{s} m_e
a_{2DEG})]^{1/2}$, where
$a_{2DEG}$ is the effective layer thickness of the Q2DEG.
Further, $\tau_\parallel$ and $\tau_\perp$
are the phenomenological longitudinal and transverse relaxation times,
respectively,
$\Omega_{10}=\Omega_{10}(n_{2DEG})=({\cal E}_1-{\cal E}_0)/\hbar$ is
the subband separation frequency of the two lowest electric subbands
of the effective confining potential (non-parabolic) of the PQW
and $f_{10}$ is the oscillator strength of the transition
$0 \rightarrow 1$.
For the PQW under consideration the subband (bottom) energies
${\cal E} _K$ ,
the envelope wave functions $\varphi_K (z)$, the Fermi energy $E_F$ and
the oscillator strengths $f_{KK'}$ are calculated self-consistently
in the framework of the Hartree approximation using the method described
in our recent paper \cite{tk31}.

Assuming monochromatic electric and magnetic fields, ${\bf E}({\bf
x},t)=
{\bf E}({\bf x},\omega)\exp(-i\omega t)$ and ${\bf H}({\bf x},t)={\bf H}
({\bf x},\omega)\exp(-i\omega t)$, respectively,  these fields are given
by the
wave equations
\begin{equation}
\vec \nabla (\vec \nabla \cdot {\bf E} ({\bf x},\omega))-
(\vec \nabla \cdot
\vec \nabla)
{\bf E} ({\bf x},\omega)={\omega^2 \over c^2}\, \varepsilon
\!\!\!\!\!\!\phantom{\varepsilon}^{\leftrightarrow}
({\bf x},\omega)\,
{\bf E}({\bf x},\omega),
\label{7}
\end{equation}
and
\begin{equation}
\vec \nabla (\vec \nabla \cdot {\bf H} ({\bf x},\omega))-
(\vec \nabla \cdot
\vec \nabla)
{\bf H}({\bf x},\omega)=-i\,\omega \,\varepsilon_0
\, \vec \nabla \times
[\varepsilon
\!\!\!\!\!\!\phantom{\varepsilon}^{\leftrightarrow}
({\bf x},\omega) {\bf E}({\bf x},\omega)] .
\label{8}
\end{equation}
We assume that the incident plane wave travels in the half-space
$z>z_0$ ($\nu=0$),
filled by vacuum, within the $x-z$ plane in the negative z-direction
with the wave vector ${\bf k}^{(0)}=
(k_x,0,-k_z^{(0)})$,  having the components
$k_x=(\omega/c)sin\Theta_0$ and
$k_z^{(0)}=(\omega/c)cos\Theta_0$.
In this case the here considered {\it
p-polarization } (TM-waves) with
\begin{equation}
{\bf E} (x,z|\omega)=[E_x (x,z|\omega),0,E_z(x,z|\omega)]
\label{9}
\end{equation}
and
\begin{equation}
\!\!\!\!\!\!\!\!\!\!\!\!\!\!\!\!\!\!\!\!\!\!\!\!\!\!\
{\bf H} (x,z|\omega)=[0,H_y (x,z|\omega),0]
\label{10}
\end{equation}
is independent from the {\it s-polarization} (TE-waves).

According to the Floquet-Bloch theorem we have in the grating region
${\bf E}^{(1)}(x+md,z|\omega)=
\exp(ik_{x}md){\bf E}^{(1)}(x,z|\omega)$, where the Bloch wave
vector component
$k_x$ is
defined in the first Brillouin zone: $-\pi/d<k_x\le \pi /d$.
Whereas in a
multilayer system of homogeneous layers ${\bf k}_\parallel =(k_x,k_y)$
is a
conserved quantity of the whole system, the periodic structure of the
grating coupler produces an infinite number of propagating waves with
${\bf k}_{\parallel n}=(k_{xn},k_y)$, where $k_{xn}=k_x+G_n$ and
$G_n={2\pi\over d}n$,
$n=0,\pm 1,\pm 2,\ldots $ is the reciprocal lattice vector. In the
presence of the grating the reflected and the transmitted beams are
represented
by {\it Fourier series (Rayleigh expansion {\rm \cite{tk33a}})}:
\begin{equation}
{\bf E}(x,z|\omega)=\sum_{n=-\infty}^{\infty} exp(ik_{xn} x)
{\bf E}_n(z,\omega)
,\,\,\, \mbox{if}\,\, z>z_0\, \,\mbox{and}\,\, z<z_1 .
\label{11}
\end{equation}
This ansatz has to fulfil the wave equation in each layer.
In the region $z>z_0$ we have
\begin{eqnarray}
&& {\bf E}^{(0)}(x,z|\omega)= \sqrt{{\varepsilon_0\over \mu_0}}
\bigl\{A_p^{(0)}
exp[i(k_x x-k_z^{(0)} z)](k_z^{(0)}{\bf e}_x+k_x{\bf e}_z)
\nonumber \\  &&
+\sum_{n=-\infty}^\infty
B_{pn}^{(0)}exp[i(k_{xn}x+k_{zn}^{(0)}z)]
(-k_{zn}^{(0)}{\bf e}_x+k_{xn}{\bf e}_z)\bigr\},
\label{12}
\end{eqnarray}
where $A_p^{(0)}$ is the field amplitude of the incident wave,
$B^{(0)}_{pn}$
is the amplitude of the $n$th diffraction-order reflected wave and
${\bf e}_\alpha$ is the unit vector along the $\alpha$-axis. From the
dispersion relation in vacuum $k_n^{(0)}=|{\bf k}_n^{(0)}|=\omega/c$ it
follows
$k_{zn}^{(0)}=[
\omega^2/c^2-k_{xn}^2]^{1/2}$. For a layer filled by an isotropic
semiconductor we have
\begin{eqnarray}
&&
{\bf E}^{(\nu)}(x,z|\omega)= \sqrt{{\varepsilon_0\over
\mu_0}}{1\over \varepsilon_\nu(\omega)}
\sum_{n=-\infty}^\infty
exp(ik_{xn} x) \{ A_{pn}^{(\nu)}
exp[-ik_{zn}^{(\nu)} (z-z_{\nu-1})]
\nonumber \\ &&
\times (k_{zn}^{(\nu)}{\bf e}_x+k_{xn}{\bf e}_z)
+B_{pn}^{(\nu)}exp[ik_{zn}^{(\nu)}(z-z_{\nu-1})](-
k_{zn}^{(\nu)}{\bf e}_x+
k_{xn}{\bf e}_z) \},
\label{13}
\end{eqnarray}
where $k_{zn}^{(\nu)}=[\varepsilon_\nu(\omega) \omega^2/c^2-
k_{xn}^2]^{1/2}$
is valid, where $A_{pn}^{(\nu)}$ is the amplitude of the nth order
diffracted wave propagating downwards and $B_{pn}^{(\nu)}$ of
that propagating upwards in layer $\nu$,
and for the anisotropic layer contaning the Q2DEG it follows
\begin{eqnarray}
&&
{\bf E}^{(\nu)}(x,z|\omega)= \sqrt{{\varepsilon_0\over
\mu_0}}{1\over \varepsilon_{xx}^{(\nu)}(\omega)}
\sum_{n=-\infty}^\infty exp(ik_{xn} x) \{ A_{pn}^{(\nu)}
exp[-ik_z^{(\nu)} (z-z_{\nu-1})]
\nonumber \\ &&
\times (k_{zn}^{(\nu)}{\bf e}_x+k_{xn}{\bf e}_z)
+B_{pn}^{(\nu)}exp[ik_{zn}^{(\nu)}(z-z_{\nu-1})](-
k_{zn}^{(\nu)}{\bf e}_x+
k_{xn}{\bf e}_z)\}
\label{14}
\end{eqnarray}
with $k_{zn}^{(\nu)}=[\varepsilon_{xx}^{(\nu)}(\omega)
\omega^2/c^2-
\varepsilon_{xx}^{(\nu)}k_{xn}^2/\varepsilon_{zz}^{(\nu)}]^{1/2}$.

In the grating region we represent the fields by the {\it modal-expansion
method} \cite{tk35,tk36},
i.e. represent the electromagnetic fields as a sum of the eigensolutions
of the wave equation in the grating layer. One solves the wave equation
in
each stripe of the grating and requires the boundary conditions
$[\![E_z]\!]
=0$ and $[\![H_y]\!]=0$, where $[\![A]\!]$ denotes the change of $A$
evaluated at the interface.
This results in the
dispersion relation of the modes
\begin{equation}
{(\varepsilon_a \beta_b )^2+(\varepsilon_b \beta_a)^2
\over 2 \varepsilon_a \varepsilon_b \beta_a \beta_b}
sin(\beta_a a)sin(\beta_b b)-cos(\beta_a a)cos(\beta_b b)
+cos(k_x d) =0.
\label{15}
\end{equation}
Equation (\ref{15}) determines for a given pair $(\omega,k_x)$ a set of
eigenvalues
$\{k_{zl}^{(1)}\}$. The electromagnetic field in the grating layer is the
sum over all eigenfunctions:
\begin{eqnarray}
&&
{\bf E}^{(1)}(x,z|\omega)={c\over \omega} \sum_l  \varepsilon_\xi\{
X_{+ l}(x)k_{zl}^{(1)} \left [ A_{l}^{(1)} e^{-ik_{zl}^{(1)}z}-
B_{l}^{(1)} e^{ik_{zl}^{(1)}z}\right ] {\bf e}_x
\nonumber \\ &&
+ X_{- l}(x)\beta_{\xi l} \left [ A_{l}^{(1)} e^{-ik_{zl}^{(1)}z}+
B_{l}^{(1)} e^{ik_{zl}^{(1)}z}\right ] {\bf e}_z \},
\label{16}
\end{eqnarray}
where
\begin{equation}
X_\pm (x)=\left \{ \matrix{ D_{al} e^{i \beta_{al} (x-md)} \pm F_{al}
e^{-i \beta_{al} (x-md)} & ; &
md<x<md+a, \cr & & \cr D_{bl} e^{i \beta_{bl} (x-md-a)} \pm
F_{bl}
e^{-i \beta_{bl} (x-md-a)} & ; &
md+a<x<(m+1)d \cr} \right .  .
\label{17}
\end{equation}
and $\beta_{\xi l}=[\varepsilon_\xi(\omega)\omega^2/c^2-
(k_{zl}^{(1)})^2]^{1/2}$.

The electromagnetic fields, given by Eqs. (\ref{12})-(\ref{14})
and (\ref{16}) have to fulfil the electromagnetic
boundary conditions $[\![E_x]\!]=0$ and $[\![H_y]\!]=0$ at the interfaces
between the different layers of the multilayer system. The most
profitable
method to represent the results is the {\it transfer-matrix method}. The
transfer matrix relates the field amplitudes in one layer to that of a
different layer. Without the grating coupler the electromagnetic field is
characterized by two amplitudes in each layer and hence, the transfer
matrix
is a $2\times 2$ matrix. But in the presence of the grating coupler the
electromagnetic field is represented in each layer by the Rayleigh
expansion
with an infinite number of field amplitudes. It is therefore necessary
for
practical calculations to restrict on a finite number of scattered
modes:
$-n_{max}\le n \le n_{max}$.
The field amplitudes of the different diffraction orders
of the transmitted and
reflected waves in layer $\nu$, we arrange in form of column matrices
($2n_{max}+1$ dimensional vectors):
\begin{eqnarray}
&&
{\cal A}_{p n_{max}}^{(\nu)}=
(A_{p 0}^{(\nu)},A_{p-1}^{(\nu)},A_{p 1}^{(\nu)},\ldots,
A_{p-n_{max}}^{(\nu)},A_{p n_{max}}^{(\nu)}),
\label{18} \\
&&
{\cal B}_{p n_{max}}^{(\nu)}=
(B_{p 0}^{(\nu)},B_{p -1}^{(\nu)},B_{p 1}^{(\nu)},\ldots,
B_{p -n_{max}}^{(\nu)},B_{p n_{max}}^{(\nu)}).
\label{19}
\end{eqnarray}
Note that the incident wave contains
only the zeroth order ${\cal
A}_{pn_{max}}^{(0)}=(A_{p0}^{(0)},0,\ldots,0)$.
The resulting matrix equation, which relates the field amplitudes in
layer $\nu$ with that in layer $\nu +1$, is given by
\begin{equation}
\left (
\begin{array}{c}
{\cal A}_{p n_{max}}^{(\nu)} \\
{\cal B}_{p n_{max}}^{(\nu)}
\end{array}
\right ) = \left (
\begin{array}{cc}
{\cal T}_{p n_{max}}^{11} & {\cal T}_{p n_{max}}^{12} \\
{\cal T}_{p n_{max}}^{21} & {\cal T}_{p n_{max}}^{22}
\end{array}
\right ) \left (
\begin{array}{c}
{\cal A}_{p n_{max}}^{(\nu+1)} \\
{\cal B}_{p n_{max}}^{(\nu+1)}
\end{array}
\right ),
\label{20}
\end{equation}
where the $2(2n_{max}+1)\times 2(2n_{max}+1)$
dimensional transfer matrix is
\begin{equation}
{\bf T}_{p n_{max}}(\nu,\nu+1)=\left (
\begin{array}{cc}
{\cal T}_{p n_{max}}^{11} & {\cal T}_{p n_{max}}^{12} \\
{\cal T}_{p n_{max}}^{21} & {\cal T}_{p n_{max}}^{22}
\end{array}
\right ).
\label{21}
\end{equation}
The submatrices are given by
\begin{equation}
{\cal T}_{p n_{max}}^{ij}=\left (
\begin{array}{ccc}
[{\cal T}_{p n_{max}}^{ij}]_{11} & \ldots &
[{\cal T}_{p n_{max}}^{ij}]_{1(2n_{max}+1)}\\
\vdots &  & \vdots \\
{[{\cal T}_{p n_{max}}^{ij}]_{(2n_{max}+1)1}} & \ldots &
[{\cal T}_{p n_{max}}^{ij}]_{(2n_{max}+1)(2n_{max}+1)}
\end{array}
\right ),
\label{22}
\end{equation}
where $i,j=1,2$. These submatrices are
$(2n_{max}+1)\times(2n_{max}+1)$ dimensional matrices,
where we have arranged the elements in the following manner:
$[{\cal T}_{p n_{max}}^{ij}]_{11}=$
$[{\cal T}_{p n_{max}}^{ij}]_{n=0n'=0}$,
$[{\cal T}_{p n_{max}}^{ij}]_{12}=$
$[{\cal T}_{p n_{max}}^{ij}]_{n=0n'=-1}$,
$[{\cal T}_{p n_{max}}^{ij}]_{21}=$
$[{\cal T}_{p n_{max}}^{ij}]_{n=1n'=0}$, $\ldots$
$[{\cal T}_{p n_{max}}^{ij}]_{(2n_{max}+1)(2n_{max}+1)}=$
\linebreak[4]
$[{\cal T}_{p n_{max}}^{ij}]_{n=n_{max}n'=n_{max}}$.
The transfer matrix can be written as
${\bf T}_{p n_{max}}(\nu,\nu+1)=$
$[{\bf P}_{p n_{max}}(\nu)]^{-1}$
$[{\bf D}_{p n_{max}}(\nu)]^{-1}$
${\bf D}_{p n_{max}}(\nu+1)$, where
${\bf P}_{p n_{max}}(\nu)$ is the {\it propagation matrix},
which describes the propagation of the diffracted waves
in layer $\nu$, and
${\bf D}_{p n_{max}}(\nu)$ is the {\it dynamical matrix}, which depends
on the polarization of the waves.
The different matrices  are derived explicitely in Ref. \cite{tk33},
where the reader can find any details.
As shown in Ref. \cite{tk33} the propagation matrix of a homogeneous
layer is given by
\begin{equation}
{\bf P}_{p n_{max}}(\nu)=\left (
\begin{array}{cc}
{\cal P}_{p n_{max}}^1(\nu) & 0 \\
0 & {\cal P}_{p n_{max}}^2(\nu)
\end{array}
\right ),
\label{23}
\end{equation}
where the elements of the submatrices read
\begin{equation}
[{\cal P}_{p n_{max}}^1(\nu)]_{ln}
=\exp (ik_{zn}^{(\nu)} d_\nu) \delta_{ln}
\label{24}
\end{equation}
and
\begin{equation}
{\cal P}_{p n_{max}}^2(\nu) =({\cal P}_{p n_{max}}^1(\nu))^{-1}.
\label{25}
\end{equation}
Herein, we have arranged the elements of the submatrices
in the following form:
${\cal P}_{11}^1={\cal P}_{n=0,l=0}^1$,
${\cal P}_{12}^1={\cal P}_{n=0,l=-1}^1$,
${\cal P}_{21}^1={\cal P}_{n=-1,l=0}^1$,\ldots,
${\cal P}_{12n_{nmax}+1}^1={\cal P}_{n=0,l=n_{max}}^1$, \ldots,
${\cal P}_{2n_{max}+12n_{max}+1}^1={\cal P}_{n=n_{max},l=n_{max}}^1$.
The dynamical matrix of a homogeneous layer is calculated to be
\begin{equation}
{\bf D}_{p n_{max}}(\nu)=\left (
\begin{array}{cc}
{\cal D}_{p n_{max}}^{11}(\nu) & {\cal D}_{p n_{max}}^{12}(\nu) \\
{\cal D}_{p n_{max}}^{21}(\nu) & {\cal D}_{p n_{max}}^{22}(\nu)
\end{array}
\right ),
\label{26}
\end{equation}
where the elements of the submatrices are given by
\begin{equation}
[{\cal D}_{p n_{max}}^{11}(\nu)]_{nl}
= [{\cal D}_{p n_{max}}^{12}(\nu)]_{nl}=\delta_{nl}
\label{27}
\end{equation}
and
\begin{equation}
[{\cal D}_{p n_{max}}^{21}(\nu)]_{nl}
=-[{\cal D}_{p n_{max}}^{22}(\nu)]_{nl}=
{k_{zn}^{(\nu)}\over \varepsilon_{xx}^{(\nu)}} \delta_{nl}.
\label{28}
\end{equation}
Equations (\ref{23}) to (\ref{28}) are valid for any optically uniaxial
medium. The corresponding expressions for isotropic media follow
if one replaces
$\varepsilon_{xx}^{(\nu)}(\omega)=\varepsilon_{zz}^{(\nu)}(\omega)=
\varepsilon_\nu(\omega)$.

Defining the transfer matrix of the whole sample by
\begin{eqnarray}
{\bf T}_{p n_{max}}^G(0,N+1)=
{\bf T}_{p n_{max}}^G(0,1)
{\bf T}_{p n_{max}}^G(1,2)
{\bf T}_{p n_{max}}(2,3)\ldots
{\bf T}_{p n_{max}}(N,N+1),
\label{a29}
\end{eqnarray}
then field amplitudes (${\cal A}_{pn_{max}}^{(0)},{\cal
B}_{pn_{max}}^{(0)}$)
in the half-space $z>z_0$ are related to that in the substrate $z<z_N$
by
\begin{equation}
\left ( \matrix{ {\cal { A}}_{pn_{max}}^{(0)} \cr & \cr {\cal {
B}}_{pn_{max}}^{(0)}
\cr } \right )={\bf T}_{p n_{max}}^G(0,N+1)\left ( \matrix{
{\cal { A}}_{pn_{max}}^{(N+1)} \cr & \cr {\cal {
B}}_{pn_{max}}^{(N+1)}\cr}
\right ) .
\label{35}
\end{equation}
Herein, the transfer matrices
${\bf T}_{p n_{max}}^G(0,1)$ und ${\bf T}_{p n_{max}}^G(1,2)$
result from the two interfaces of the grating with the homogeneous
media.
Further, the following is valid: The submatrices which result from the
interface between two homogeneous media are diagonal, whereas those
resulting from the interfaces between the grating and homogeneous
media are non-diagonal. Also the last one can be represented in the form
${\bf T}_{p n_{max}}^G(0,1)=[{\bf P}_{p n_{max}}(0)]^{-1}
[{\bf D}_{p n_{max}}(0)]^{-1}{\bf D}_{p n_{max}}^G(1)$ und
${\bf T}_{p n_{max}}^G(1,2)=[{\bf P}_{p n_{max}}^G(1)]^{-1}
[{\bf D}_{p n_{max}}^G(1)]^{-1}{\bf D}_{p n_{max}}(2)$.
The matrices $[{\bf P}_{p n_{max}}(0)]^{-1}$ and
$[{\bf D}_{p n_{max}}(0)]^{-1}$ are the inverse of the propagation
matrix and of the dynamical matrix of the half-space $z>z_0$ ($\nu=0$)
given by Eqs. (\ref{23}) and (\ref{26}), respectively, and
${\bf D}_{p n_{max}}(2)$ is the propagation matrix of layer $\nu=2$
given by Eq. (\ref{26}). In the grating region itselfs the propagation
matrix ${\bf P}^G_{p n_{max}}(0)$ has the same form as given in
Eqs. (\ref{23}) to (\ref{25}). But the dynamical matrix in the grating
region is different from that describing homogeneous layers.
The dynamical matrix of the grating is given by
\begin{equation}
{\bf D}^G_{p n_{max}}(\nu)=\left (
\begin{array}{cc}
{\cal D}_{p n_{max}}^{G,11}(\nu) &
{\cal D}_{p n_{max}}^{G,12}(\nu) \\
{\cal D}_{p n_{max}}^{G,21}(\nu) &
{\cal D}_{p n_{max}}^{G,22}(\nu)
\end{array}
\right ),
\label{30}
\end{equation}
with
\begin{equation}
[{\cal D}_{p n_{max}}^{G,11}(\nu)]_{nl}
= [{\cal D}_{p n_{max}}^{G,12}(\nu)]_{nl}=
\Gamma_{ln}^1
\label{31}
\end{equation}
and
\begin{equation}
[{\cal D}_{p n_{max}}^{G,21}(\nu)]_{nl}
=-[{\cal D}_{p n_{max}}^{G,22}(\nu)]_{nl}=
\Gamma_{nl}^2k_{zl}^{(\nu)}
.
\label{32}
\end{equation}
The different element of these matrices are
\begin{eqnarray}
\Gamma_{ln}^1&=&{1\over d} \int_0^d dx \, \exp(-ik_{xn}x)X_{+l}(x) =
\nonumber \\
&=&iD_{al}{\{1-\exp[ia(\beta_{al}-k_{xn})]\} \over \beta_{al}-k_{xn}}-
iF_{al}{\{1-\exp[-ia(\beta_{al}+k_{xn})]\} \over \beta_{al}+k_{xn}}+
\nonumber \\
&+&iD_{bl}{\{1-\exp[ib(\beta_{bl}-k_{xn})]\} \over \exp(ik_{xn}a)
(\beta_{bl}-k_{xn})}-
iF_{bl}{\{1-\exp[-ib(\beta_{bl}+k_{xn})]\} \over \exp(ik_{xn}a)
(\beta_{bl}+k_{xn})},
\label{33} \\
\Gamma_{ln}^2&=& {1\over d}
\int_0^d dx \, \exp(-ik_{xn}x){X_{+l}(x)\over \varepsilon_\xi}=
\nonumber \\
&=&iD_{al}{\{1-\exp[ia(\beta_{al}-k_{xn})]\} \over \varepsilon_a
(\beta_{al}-k_{xn})}-
iF_{al}{\{1-\exp[-ia(\beta_{al}+k_{xn})]\} \over \varepsilon_a
(\beta_{al}+k_{xn})}+
\nonumber \\
&+&iD_{bl}{\{1-\exp[ib(\beta_{bl}-k_{xn})]\} \over \exp(ik_{xn}a)
\varepsilon_b(\beta_{bl}-k_{xn})}-
iF_{bl}\!{\{1-\exp[-ib(\beta_{bl}+k_{xn})]\} \over \exp(ik_{xn}a)
\varepsilon_b(
\beta_{bl}+k_{xn})}.
\label{34}
\end{eqnarray}
Herein, the matrix elements $\Gamma_{ln}^1$ ($n=0,-1,1,...n_{max}$,
$l=1,...2n_{max}+1$) are arranged in the form:
$\Gamma_{11}^1 = \Gamma_{l=1,n=0}^1$,
$\Gamma_{12}^1 = \Gamma_{l=1,n=-1}^1$,
$\Gamma_{21}^1 = \Gamma_{l=2,n=0}^1$, $\ldots$,
$\Gamma_{1 2n_{max}+1}^1 = \Gamma_{l=1,n=n_{max}}^1$, $\ldots$,
$\Gamma_{2n_{max}+1 2n_{max}+1}^1 = \Gamma_{l=2n_{max}+1,n=n_{max}}^1$.

In the absence of the grating the here presented formalism reduces to
the
well-known $2 \times 2$ transfer-matrix method, i.e. each submatrix of
the
transfer matrix given in Eq. (\ref{20}) becomes a single complex
function
or number and the field amplitudes, Eqs. (\ref{18}) and (\ref{19}),
reduce to only one
(that of the zeroth order), respectively:
\begin{equation}
\left ( \matrix{ {{ A}}_{p 0}^{(\nu)} \cr & \cr {{
B}}_{p 0}^{(\nu)}
\cr } \right )={\bf T}_{p 0}(\nu,\nu+1)\left ( \matrix{
{{ A}}_{p 0}^{(\nu+1)} \cr & \cr { {
B}}_{p 0}^{(\nu+1)}\cr}
\right ) .
\label{b35}
\end{equation}
with
\begin{eqnarray}
\!\!{\bf T}_{p 0}(\nu,\nu +1)={1 \over 2} \left ( \matrix{
\displaystyle{[1+{\varepsilon_{xx}^{(\nu)}k_z^{(\nu+1)}\over
\varepsilon_{xx}^{(\nu+1)}
k_z^{(\nu)}}]exp(-ik_z^{(\nu)}d_\nu)} &
\displaystyle{[1-{\varepsilon_{xx}^{(\nu)}k_z^{(\nu+1)}\over
\varepsilon_{xx}^{(\nu+1)}
k_z^{(\nu)}}]exp(-ik_z^{(\nu)}d_\nu)} \cr & \cr  & \cr
\displaystyle{[1-{\varepsilon_{xx}^{(\nu)}k_z^{(\nu+1)}\over
\varepsilon_{xx}^{(\nu+1)}
k_z^{(\nu)}}]exp(ik_z^{(\nu)}d_\nu)} &
\displaystyle{[1+{\varepsilon_{xx}^{(\nu)}k_z^{(\nu+1)}\over
\varepsilon_{xx}^{(\nu+1)}
k_z^{(\nu)}}]exp(ik_z^{(\nu)}d_\nu) }\cr } \right ).
\label{c35}
\end{eqnarray}

For a multilayer system with a grating coupler, however,
transmitted waves of the order
$n=0,\pm 1,\pm 2, \ldots \pm n_{max}$ occur.
The quantity measured in the experiments is the {\it time-averaged power
transmission coefficient}, which is calculated to be
\begin{equation}
T_p= {1\over 2 \varepsilon_{s N+1} |A_{p0}^{(0)}|^2
k_{z0}^{(0)}}
\sum_{n,n'=-n_{max}}^{n_{max}}
A_{pn}^{(N+1)*}A_{pn'}^{(N+1)}e^{i\phi_{nn'}(x,z)}
(k_{zn}^{(N+1)*}+k_{zn'}^{(N+1)}),
\label{36}
\end{equation}
where the symbol $*$ means complex conjugate and
\begin{equation}
\phi_{nn'}(x,z)=x(k_{xn'}-k_{xn})+(z-z_N)(k_{zn}^{(N+1)*}-
k_{zn'}^{(N+1)}).
\label{37}
\end{equation}
The transmission coefficient depends on both $x$ and $z$ because the
grating
produces an diffraction pattern along the $x$-direction. It is important
to note that
not all the orders of diffracted waves can propagate through the sample
and
from the the sample surface to the detector. Only
those waves for which $k_{zn}^{(0)}$ is real and positive, i.e. if
${\omega^2\over c^2}
>({\omega\over c}sin\Theta_0+{2\pi \over d}n)^2$ are \it propagating
\rm
waves above and below the sample.
In the case that $k_{zn}^{(0)}$ is pure imaginary, i.e. if
${\omega^2\over c^2}
<({\omega\over c}sin\Theta_0+{2\pi \over d}n)^2$ the corresponding
wave is an
\it evanescent \rm wave with the decay length $L_n=(2|Im
k_{zn}^{(0)}|)$ and thus, exists only in the near-field of the grating.
Because in the experiment the distance between the sample and the
detector is
usually much larger than the decay lengths of the evanescent waves,
which
cannot transport energy in the negative $z$-direction if the medium
below
the sample is non-absorbing, only
the propagating waves are detected. But it is important that the
evanescent waves couple with the collective excitations
of the Q2DEG inside the
PQW, which usually is in the near
vicinity of the grating, leading to a resonant decrease of the
reflected and transmitted radiation.
Hence, the measured transmission coefficient
contains via the boundary conditions at the interfaces of the
grating with the
homogeneous layers, which couple the different diffracted orders,
all information about the evanescent higher-order diffracted waves.
In the typical situation only the zero-order diffracted wave
is propagating in the
sourrounding. If in this case, the influence of the evanescent waves
below the sample on the
transmission coefficient of Eq. (\ref{37}) can be neglected. Then,
the transmission
coefficient reads
\begin{equation}
T_p={|A_{p0}^{(N+1)}|^2k_{z0}^{(N+1)} \over \varepsilon_{s
N+1}
|A_{p0}^{(0)}|^2
k_{z0}^{(0)}}.
\label{38}
\end{equation}
With increasing period $d$ of the grating it becomes possible that the
first-order diffracted waves become propagating. In the vacuum above
and below
the sample this takes place at $d=\lambda_0$, where $\lambda_0=2\pi /
{k^{(0)}}= 2 \pi c/ \omega$ is the vacuum wavelength at the
frequency of the ISR. If this becomes true
the transmission spectrum as well as the reflection spectrum
show the so-called {\it Rayleigh anomaly}
\cite{tk33a,tk33b,tk38,tk37,tk37b}. It consists on a rapid variation
in the amplitudes of the diffracted orders corresponding to the onset
(evanescent $\to$ propagating) or disappearance (propagating $\to$
evanescent) of a particular diffracted order. This is true because
with the appearance of a new order of diffracted wave a rearrangement
of the field amplitudes of the other propagating diffracted waves
is caused. The wavelength were this takes place is called {\it
Rayleigh wavelength} $\lambda_R$. But in general, {\it two types
of anomalous effects, called Wood's anomalies} \cite{tk39}, occur:
(i) the above described Rayleigh wavelength type and (ii) the
resonance-type anomaly. The second type of Wood's anomalies is
connected with the excitation of a leaky surface wave propagating
along the metallic grating \cite{tk38,tk37b,tk40}.

If $2n_1+1$ waves are propagating below the sample (zero order plus
forward and backward diffracted waves of higher order),
valid if $d \ge n_1 \lambda_0$, it depends on the detector if the
diffraction pattern is measured or the spatial average.
Usually the last one is
realized and then the transmission coefficient, which only includes the
action of the propagating waves below the sample, reads
\begin{equation}
T_p=
{1\over \varepsilon_{s N+1}|A_{p0}^{(0)}|^2 k_{z0}^{(0)}}
\sum_{n=-n_1}^{n=n_1}|A_{pn}^{(N+1)}|^2k_{zn}^{(N+1)}.
\label{39}
\end{equation}

The experimental and theoretical investigations performed in this work
are focused on the question about the \it coupling efficiency \rm of the
grating to excite the ISR of the Q2DEG
synthesized in the PQW with one occupied subband for a given
semiconductor heterostructure. It is known (for details see
\cite{tk33}) that the coupling efficiency
depends on the one hand side on the grating parameters (period, height,
mark-to-space ratio, material) but on the other hand side on the
properties
of the multilayer system (distance between grating and Q2DEG and
between
the Q2DEG and the lower boundary of the sample, material, etc.). Here,
we
investigate in detail the coupling efficiency of FIR transmission in
dependence on the grating
period $d$ and all the other parameters of the the sample remain fixed.

\section{Numerical Results and Discussion}
The numerical calculations are done for the six-layer system depicted in
Fig. 3 with the following material parameters:
layer $\nu=1$ is a $Ag$ grating layer with $\omega_{p_a}=
5.69\times 10^{15} s^{-1}$ and $\gamma_a=7.596 \times 10^{13} s^{-1}$ of
height
$d_1=h=50nm$; layer
$\nu=2$ is a $GaAs$ layer of thickness $d_2=10nm$ and
$\varepsilon_{s2}=12.87$ and layer $\nu=3$ is a
$Ga_{1-x}Al_xAs$ ($x=0.25$) layer with
$\varepsilon_{s3}=12.21$ and
of thickness $d_3=241nm$;
the fourth layer contains the PQW with the bare confining energy
$\hbar \Omega=11meV$, in which the Q2DEG is synthesized. For simplicity
we
assume for the background dielectric constant of this layer a
homogeneous dielectric constant with
the parameters of
$GaAs$:
$\varepsilon_{s4}=12.87$
and $n_{2DEG}=2\times 10^{11} cm^{-2}$,
$\tau_\parallel=1\times 10^{-11}s$,
$\tau_\perp=1\times 10^{-12}s$ and $d_4=a_{2DEG}=18.7
nm$.
Layer $\nu=5$ is $224nm$ thick and
consists of $Ga_{1-x}Al_xAs$ and the layer $\nu=6$ is a $500nm$
thick $GaAs$ layer.
In the numerical calculations the substrate (region $\nu=7$) is
filled with $GaAs$ and it is assumed to be of infinite
thickness in difference to the experimental situation.
In the experimental sample the substrate is wedged with a mean
thickness of
approximatly $500 \mu m$. In both, theory and experiment this is done
to avoid the Fabry-Perot resonances in the substrate.

The numerically calculated
relative transmission coefficient $-\Delta T/T$ is
plotted in Fig. 4 for different periods $d$ of the grating and fixed
mark-to-space ratio t=1.
The wave vector transfer from the incident wave to the ISR is for
$d=4\mu m$: $k_{x_1}=1.57\times 10^4 cm^{-1}$,
$k_{x_2}=3.14\times 10^4cm^{-1}, \ldots$;
$d=6\mu m$: $k_{x_1}=1.047\times 10^4 cm^{-1}$,
$k_{x_2}=2.094\times 10^4cm^{-1}, \ldots$;
$d=10\mu m$: $k_{x_1}=6.238\times 10^3 cm^{-1}$,
$k_{x_2}=1.256\times 10^4cm^{-1}, \ldots$;
$d=20\mu m$: $k_{x_1}=3.14\times 10^3 cm^{-1}$,
$k_{x_2}=6.28\times 10^3cm^{-1}, \ldots$;
$d=30\mu m$: $k_{x_1}=2.094\times 10^3 cm^{-1}$,
$k_{x_2}=4.188\times 10^3cm^{-1}, \ldots$;
and for
$d=40\mu m$: $k_{x_1}=1.57\times 10^3 cm^{-1}$,
$k_{x_2}=3.14\times 10^3cm^{-1}, \ldots$.
It becomes obvious from both experiment (Fig. 2) and theory (Fig. 4)
that the efficiency to excite the ISR of a
Q2DEG inside a given multilayered-quantum-well system increases
with increasing
period $d$ of the grating coupler. This is true for the given
parameters up to
$d=10 \mu m$ in excellent agreement between theory and experiment.
It is seen that analogous to the experimental result, the theoretically
calculated line shapes of the ISR peak show at smaller periods a
Lorentzian form, which becomes more and more asymmetric with
increasing grating period.
For $d=20 \mu m$ we have the largest magnitude of the maximum in
$-\Delta T/T$, but with an asymmetric shape. Increasing further the
period
transforms the maximum to a minimum which vanishes for $d \to
\infty$. Please note that in the case of the asymmetric line shape
$T(n_{2DEG})>T(0)$ is valid, i.e. more light is transmitted through
the sample with the Q2DEG.

Because it is difficult to define something like the coupling
efficiency for a
strongly disturbed line shape, we choose the integrated relative
transmission
$  \Gamma {}$  of the detected lines as a measure for it. The result is
shown in Fig. 5, where  we plot the integrated relative transmission
of the
resonant structures for the measured and calculated spectra as a
function of the grating period.
For $d <  \lambda {}_{ sub }$ we  observe a steady increase of  $\Gamma
{} $
with increasing period $d$. Beyond the vertical dashed line which
indicates the condition
$d =  \lambda {}_{ sub }$  the integrated
relative transmission decreases very
rapidly. It should be noted, however, that our experimental setup only
integrates over a finite solid angle given by the distance between the
sample and the waveguide and by the waveguides aperture. This fact
might be important especially for propagating higher order waves as
we point out below.

In the case considered here for $d<17.5\mu m$ only the zeroth-order
diffracted
wave is a propagating wave in vacuum and in the $GaAs$ substrate.
But for a grating
period of $d=17.5 \mu m$ at
$\omega=3\times 10^{13}s^{-1}$ ($\bar \nu= 159.3 cm^{-1}$) the
first-order diffracted wave additionally becomes propagating in the
$GaAs$ substrate and
for $d=62.83 \mu m$ this becomes also true in the region filled by
vacuum.
In the framework of the $\varepsilon_s$-approximation the onset of the
propagation of the first-order diffracted wave in the GaAs substrate,
which is effectively formed by the regions $\nu=6$ and $7$, is
just at $d=\lambda _{sub}$, where $\lambda _{sub}= 2 \pi c /(\sqrt
{\varepsilon_
{s6}} \omega)$ is the wavelength of the FIR light in the substrate
at the frequency of the ISR. In this case it follows from
$k_{z1}^{(6)}=0$ that $E_{x1}^{(6)}=0$ and thus, the electric field
of this wave is polarized pure perpendicular to the interfaces
in the substrate and in all the other layers filled with $GaAs$.
The physical situation inside the PQW is very similar but becomes
complicated
due to the resonance structure of the complex dielectric tensor,
which may vary
strongly in the near vicinity of the frequency of the ISR.
The onset of the propagation of the first-order diffracted wave
in the substrate is seen in
the calculated relative transmission coefficient.
The calculated curves show for $d=10 \mu m$,
$d=20 \mu m$ and $d=30 \mu m$ the Rayleigh anomaly at $\omega=5.3\times
10^{13} s^{-1}$ ($\bar \nu = 281.5 cm^{-1}$),
$ \omega=2.6 \times 10^{13} s^{-1}$
($\bar \nu = 138.1 cm^{-1}$) and $\omega = 1.7 \times 10^{13} s^{-1}$
($\bar \nu = 90.3 cm^{-1}$), respectively.
It becomes obvious from the theoretical and
experimental $-\Delta T/T$ curves that just under these conditions
the transmission peaks begin to deform.
For conditions below the onset of the Rayleigh anomaly one measures
only the
zeroth-order beam. But the peak in the relative transmission spectrum
results from the evanescent higher-order beams which couple with the
{\it non-radiative} intersubband plasmons. This situation is changed
above the onset of the Rayleigh anomaly, where the first-order
diffracted beam couples with the {\it radiative} ISR. Because the
collective intersubband excitations are
nearly dispersionless in the long-wavelength
limit, the resulting ISR peak is a sum of the action of all diffracted
beams with $k_{x1}, k_{x2}, k_{x3},\ldots$.
Our detailed numerical analysis shows that for the infinite thick
$GaAs$ substrate only the Rayleigh wavelength type
anomaly is responsible for the deformation of the peak. The
asymmetric line shape is thus caused by the superimposition of the
ISR peak with the Rayleigh anomaly. For the chosen system the
resonance-type anomaly is absent in the plotted frequency range, i.e.
the comparison of the transmission $T(0)$ and $T(n_{2DEG})$ gives no
indication for the excitation of a leaky surface wave carried by the
grating.
It is noticeable that the diffracted peaks resulting from the
excitation of the intrasubband plasmon
$\omega _p^{00}(q_{\parallel})$ at $ q_{\parallel} =k_{x1},
k_{x2}, k_{x3},\ldots$ are always symmetrically shaped.

If one would calculate $-\Delta T/T$ for a sample with $GaAs$
substrate of {\it finite} thickness, at these frequencies the
first-order diffracted wave becomes
propagating in the $GaAs$ layers of the sample and slightly below
these frequencies in the $Ga_{1-x}Al_x As$ layers. Possible
anomalies in this case occurring in the optical spectra due the
Rayleigh wavelength effect we call {\it internal Rayleigh anomalies}.
However, in this frequency range the first-order diffracted wave
is an evanescent wave in the vacuum below and above the sample.
To answer the question whether these internal Rayleigh anomalies
are the cause for the asymmetry in the experimentally detected
line shape or not, we investigated the multilayer system of
Fig. 3, but assumed the region $\nu=7$ to be filled by vacuum.
The relative transmission coefficient $-\Delta T/T$ is plotted
in Fig. 6 for two different thicknesses $d_6$, assuming here a height
of the grating of $h=10nm$. It is to be seen from Fig. 6 $(a)$ that
for $d_6=4\mu m$ the Lorentzian shaped maximum, which appears for
$d=4nm$, deforms with increasing grating period, quite similar as
shown in Fig. 4, it becomes asymmetrically shaped and for $d=40\mu m$
the maximum is transformed in a minimum. It is to be seen from the
spectra in Fig. 6 $(a)$, calculated for $d=20 \mu m$ and $d=30 \mu m$,
that the internal Rayleigh anomalies are absent and thus, they have
no significant influence on $-\Delta T/T$. Here, the question arises
about the mechanism which is responsible for the pronounced asymmetric
profiles. In Fig. 6 $(b)$ we plotted the corresponding relative
transmission spectra for $d_6=500nm$. It becomes obvious that in this
case the best coupling efficiency is obtained for $d=6\mu m$.
For larger
grating periods the peak height decreases but with a nearly stable
line shape. It should be noted that whereas the exponential function
of the first-order diffracted wave, appearing in the propagation matrix
of layer $\nu =6$, Eq. (\ref{24}), is nearly one for small $d_6$ in
the plotted frequency range, it varies rapidly with the frequency
for larger $d_6$. Such a rapid variation may cause Fabry-Perot-like
resonances of the first-order diffracted wave in the multilayer
system. Further, we have calculated the total power absorption
$A_p(n_{2DEG})=1-T_p(n_{2DEG})-R_p(n_{2DEG})$ for the two
configurations of Fig. 6 assuming a grating period of $d=3\mu m$. From
the comparison of $A_p(n_{2DEG})$ and $A_p(0)$ (see Fig. 7) it
becomes obvious that the $A_p(0)$ spectrum shows for $d_6=4\mu m$
a maximum which we attribute to the excitation of a leaky surface
wave in the Ag grating. This resonance superimposes the ISR resonance
to give the asymmetric line shape of the peak. For the smaller layer
thickness, $d_6=500nm$, the resonance associated with the surface wave
is absent in the plotted frequency range. Thus, we attribute the
asymmetric line shape for the finite width substrate due to the
resonance-type Wood's anomaly. It should be noted that similar line
shapes of the ISR resonance were observed in Ref. \cite{tk14} for
metal-oxide-semiconductor structures of silicon. In this paper
the asymmetric line shape is assumed to be caused by the
superimposition of the ISR with a {\it non-depolarization-shifted}
ISR. Our rigorous grating-coupler theory, however, shows that the ISR
is always accompanied by a depolarization shift which cannot be
"screened" by the grating. Thus, it seems that also in this early
experiment Wood's anomalies could be responsible for the asymmetric
line shape.

\section{Summary}

In summary, we have measured and calculated the optical response of the
ISR of a quasi two-
dimensional electron system in a PQW  with
different grating couplers on top of this structure. Using the
transfer-matrix
method of local optics and the modal-expansions method to calculate the
influence of the lamellar grating on the electromagnetic fields, the
calculated relative transmission describes very well the measured
spectrum.
It is shown that the coupling efficiency of the grating coupler to excite
the
ISR increases with increasing period of the grating up to a certain value
where the absorption peak starts to deform from a Lorentzian shape to
an asymmetric shape. This asymmetric line shape is caused by
both types of Wood's anomalies, i.e. due to the propagating
higher-order diffracted waves in the sample beyond
the threshold of the Rayleigh anomaly and due the resonance-type
anomaly (excitation of a leaky surface wave in the grating region).
In this case one excites
both radiative and non-radiative collective intersubband excitations.

\section{Acknowledgments}

We gratefully acknowledge many fruitful discussions with J. P. Kotthaus.
This work has been financially supported in part by the Deutsche
Forschungsgemeinschaft (Project No. We 1532/3-2) and the PHANTOMS network,
in part by the
Air Force Office of Scientific Research under grant AFOSR-91-0214,
by the NSF Science and Technology Center for Quantized Electronic
Structures (QUEST) DMR-91-20007, by  the NSF DMR-90-02291,
and the ONR N000-14-92-J1452. The M\"unchen - Santa Barbara cooperation
has been supported by a joint NSF - European grant (EC-US
015:9826).

\newpage

\noindent
\begin{description}
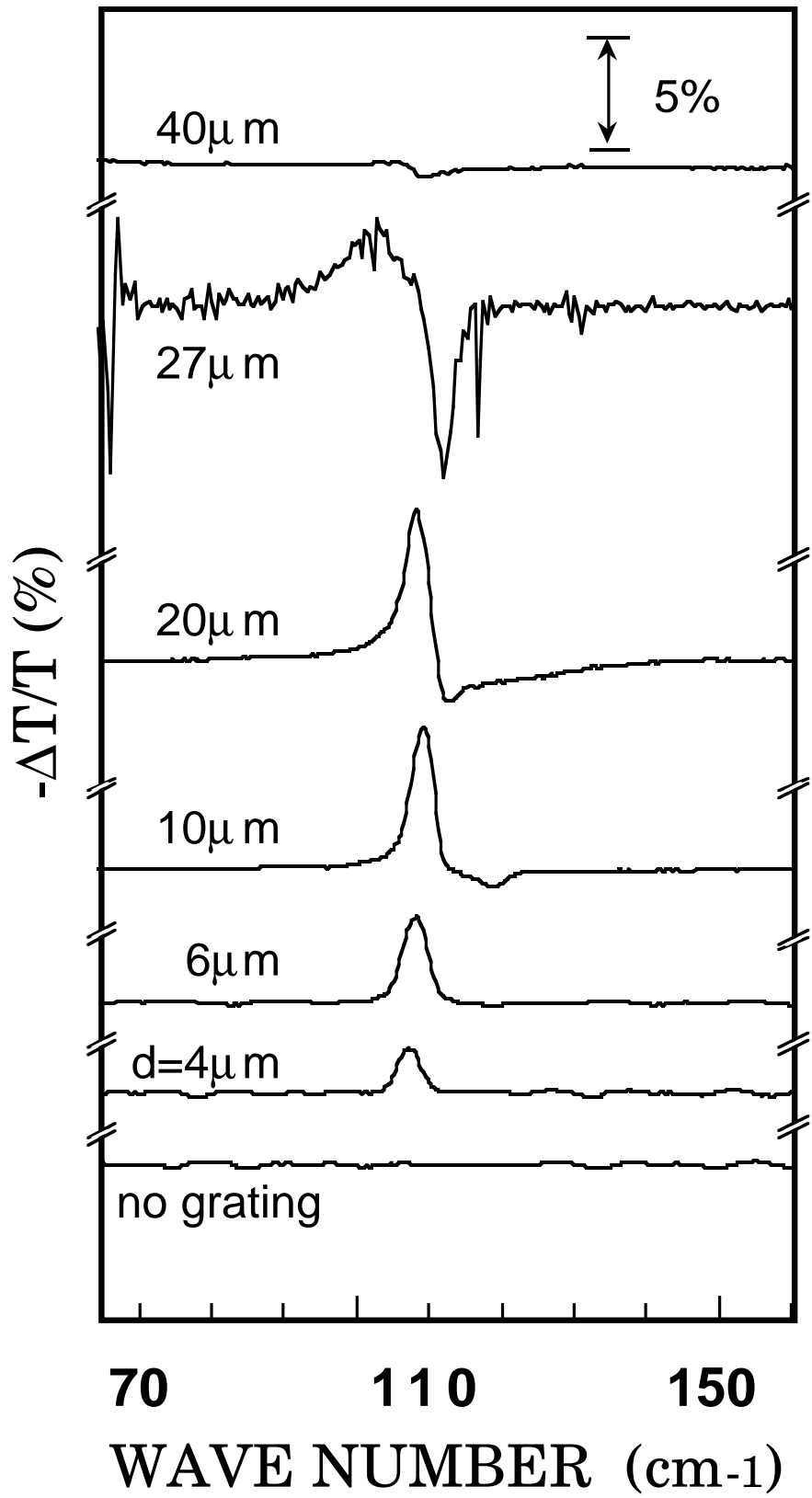
\item {Fig. 1 }
 Schematic arrangement of the layer structure of the
sample containing the PQW used in the experiments.
\end{description}
\par

\begin{description}
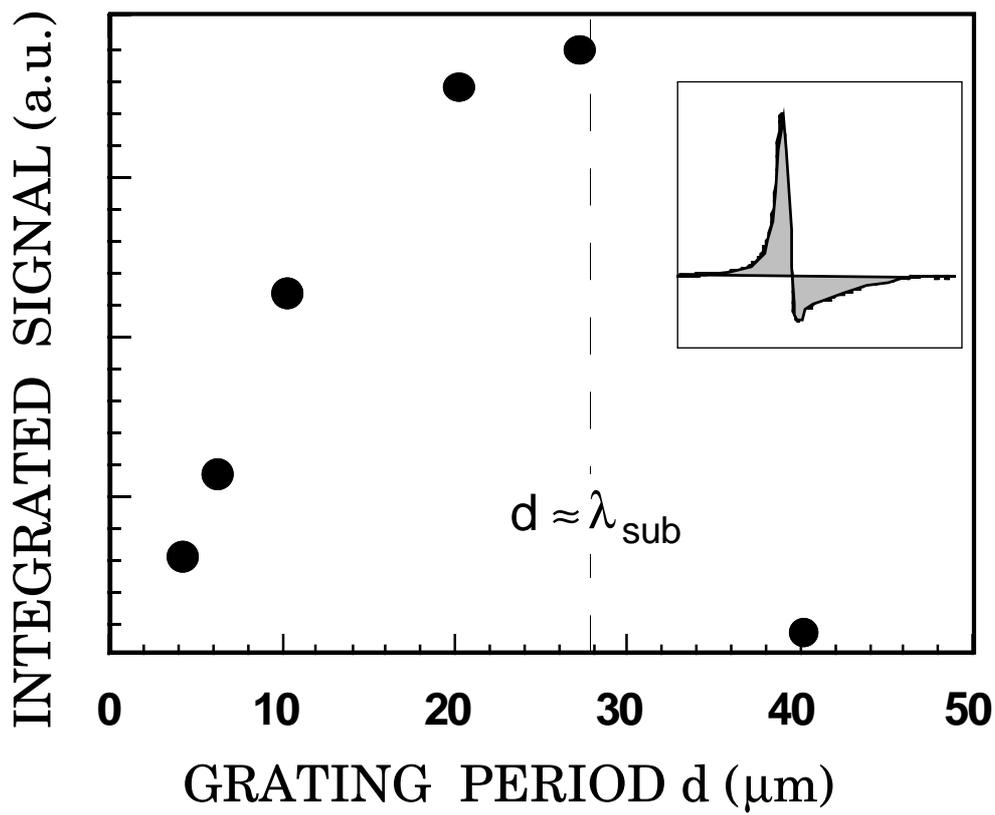
\item {Fig. 2.}
 Experimentally obtained relative transmission spectra for the ISR
in a PQW using grating couplers of different grating periods.
With increasing grating period first the height of the occurring peak
becomes larger, then the lines become strongly asymmetric and for large
grating periods the relative transmission nearly
vanishes.\end{description}
\par

\begin{description}
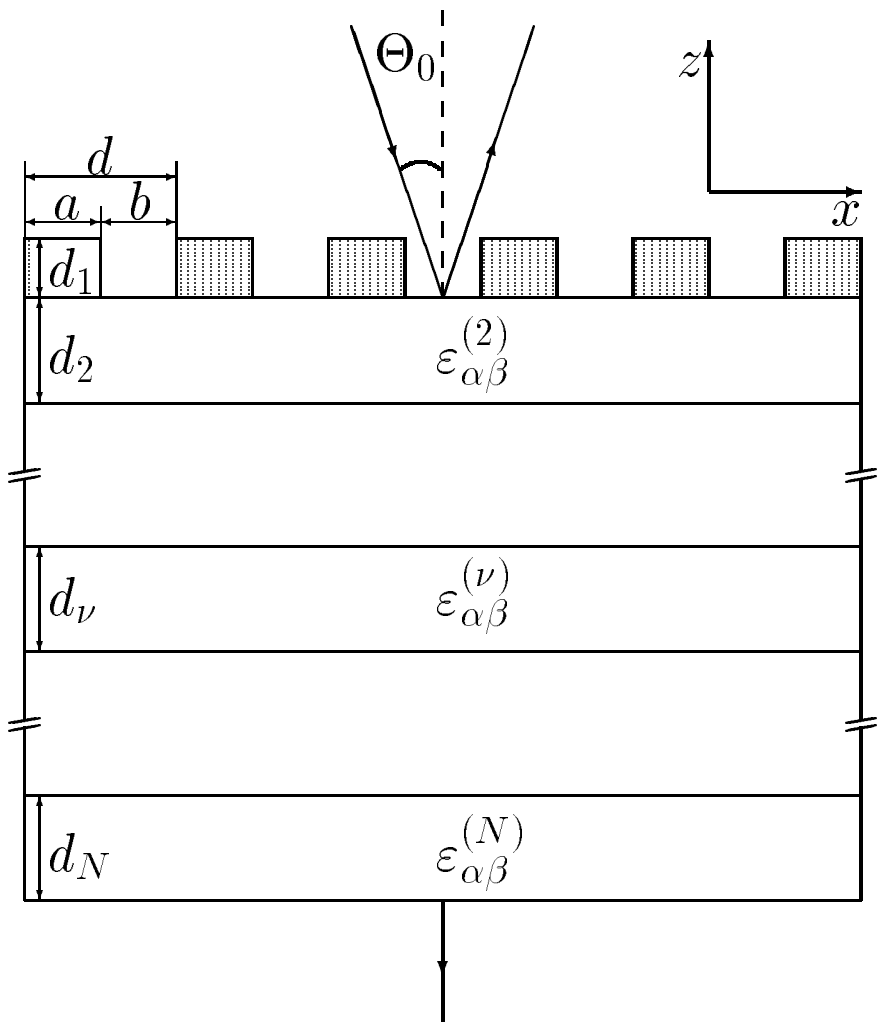
\item {Fig. 3.}
 Schematic arrangement of the geometry of the multilayer system
with grating used in the theoretical calculations.
\end{description}

\noindent
\begin{description}
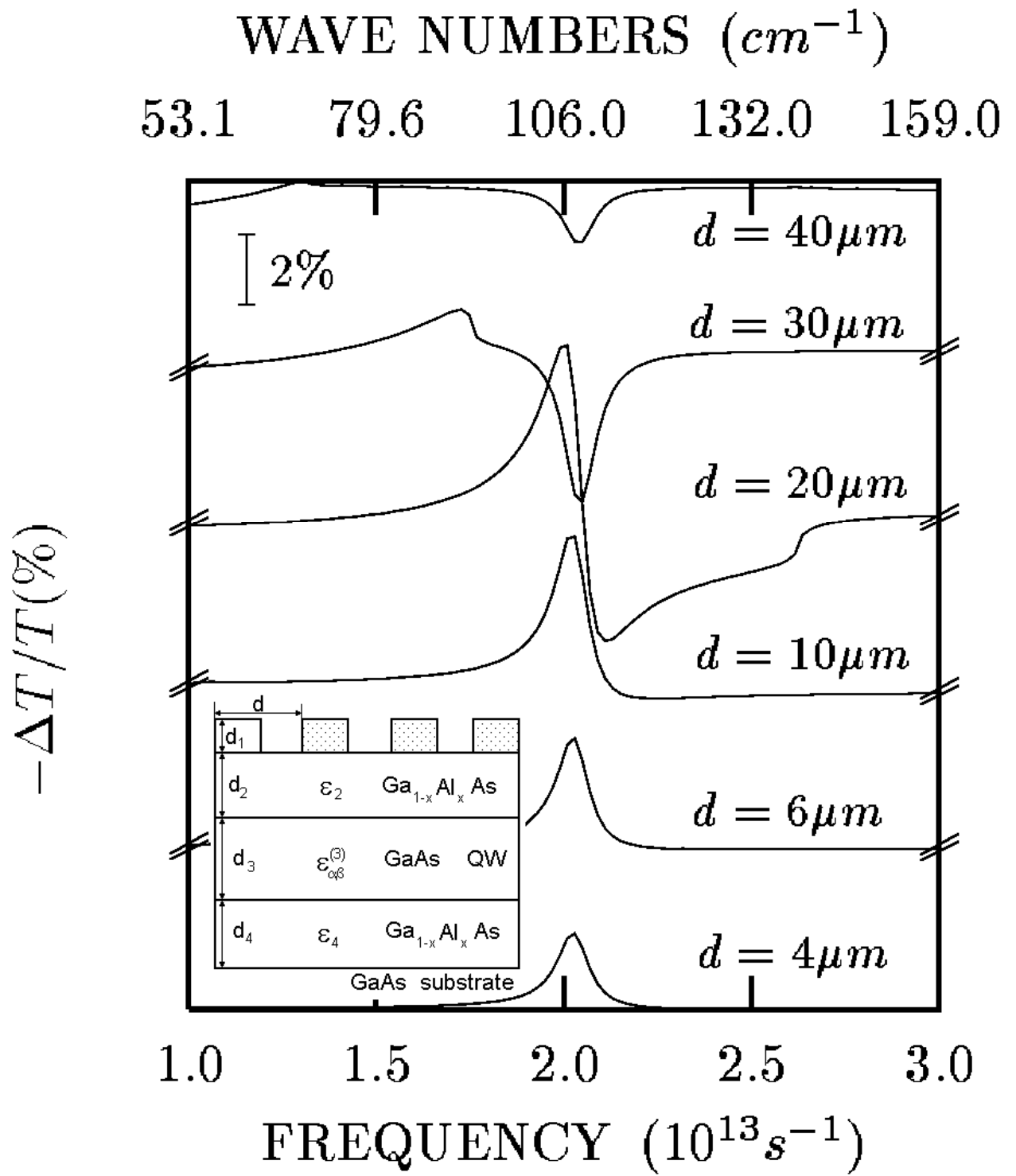
\item {Fig. 4.}
 Relative transmission coefficient $-\Delta T/T$
of the multilayer system in the near vicinity of the
ISR of the Q2DEG for different periods $d$ of the grating assuming
the region $\nu =7$ to be filled by $GaAs$.
\end{description}

\begin{description}
\item {Fig. 5.}
 Integrated relative transmission of the spectra shown
in Fig. 2 and 4. As long as the
grating period $d$ is smaller than the wavelength
$  \lambda {}_{ sub }$ of the light corresponding to the intersubband
transition, the integrated relative transmission increases. For $d$
$>  \lambda {}_{ sub}$  the integrated relative transmission rapidly
decreases.\end{description}

\begin{description}
\item {Fig. 6.}
 Relative transmission coefficient $-\Delta T/T$
of the multilayer system in the near vicinity of the
ISR of the Q2DEG for different periods $d$ of the grating of height
$h=10nm$, assuming
the region $\nu =7$ to be filled by vacuum:
(a) $d_6=4\mu m$ and
(b) $d_6=500nm$.
\end{description}

\begin{description}
\item {Fig. 7.}
 Total absorption coefficient $A(n_{2DEG})$ and $A(0)$ for the
multilayer system of Fig. 6:
(a) $d_6=4\mu m$ and
(b) $d_6=500nm$ calculated for a grating period of $d=30 \mu m$.
\end{description}

\end{document}